\documentclass[aps,preprint,showpacs,showkeys,preprintnumbers,amsmath,amssymb]{revtex4}

\usepackage{dcolumn}
\usepackage{mathrsfs}
\usepackage{bm}
\usepackage{amsmath,amssymb,epsfig,float}

\begin{document}

\title{Measurement-Induced-Nonlocality via the Unruh effect}

\author{Zehua Tian and Jiliang Jing\footnote{Corresponding author, Email: jljing@hunnu.edu.cn}}
\affiliation{Department of Physics, and Key Laboratory of Low
Dimensional Quantum Structures and Quantum
Control of Ministry of Education,\\
Hunan Normal University, Changsha, Hunan 410081, China}

\vspace*{0.2cm}
\begin{abstract}
\vspace*{0.2cm}

Treated beyond the single-mode approximation, Measurement-Induced-Nonlocality (MIN) is investigated for both Dirac and Bosonic fields in non-inertial frames. Two distinctly differences between the Dirac and Bosonic fields are: (i) the MIN for Dirac fields persists for any acceleration, while the quantity for Bosonic fields does decay to zero in the infinite acceleration limit; (ii) the dynamic behaviors of the MIN for Dirac fields is quite different from the Bosonic fields case. Besides, we also study the nonlocality for Dirac fields and find that the MIN is more general than the quantum nonlocality related to violation of Bell's inequalities. Meanwhile some discussions of geometric discord are presented too.

\end{abstract}

\vspace*{1.5cm}

\maketitle

\section{introduction}

The investigation of relativistic quantum information not only
supplies the gap of interdiscipline refer to quantum information and
relativity theory, but also has a positive promotion on the
development of them. As a result of that, this domain has been paid
much attention in the last decade \cite{Alsing1,Alsing2,Fuentes,Yi
Ling,Shahpoor,Pan,Montero2,Luoko,Montero,Montero1,Wang,Datta,Han,David,Mann}. Among them,
most papers have focused on quantum resource, e. g., quantum
entanglement \cite{Alsing1,Alsing2,Fuentes,Yi
Ling,Shahpoor,Pan,Montero,Mann} and discord \cite{Wang,Datta},
because the quantum resource plays an very important role in the quantum
information tasks such as teleportation \cite{Bennett} and
computation \cite{Horodecki,Bouwmeester}, and studying it in a
relativistic setting is very closely related to the implementation
of quantum tasks with observers in arbitrary relative motion. In
addition, extending this work to the black hole background is very
helpful for us to understand the entropy and paradox
\cite{Lee,Hawking} of the black hole.

Despite much effort has been paid to extend quantum information
theory to the relativistic setting, most papers are limited to
entanglement and discord, another foundation of quantum
mechanics--nonlocality is barely considered. Recently, Nicolai Friis
$\emph{et al}$ firstly studied the nonlocality in the noninertial
frame, and they pointed out that residual entanglement of
accelerated fermions is not nonlocal \cite{Nicolai}. Following them,
Alexander Smith $\emph{et al}$ studied the tripartite nonlocality in
the noninertial frames \cite{Smith}, and DaeKil Park considered
tripartite entanglement-dependence of tripartite nonlocality
\cite{Park}. Generally, most researchers analyzed the quantum
nonlocality by means of Bell's inequalities \cite{J. F. Clauser} for
bipartite system and Svetlichny inequality for tripartite system
\cite{G. Syetlichny}, respectively. Because, these inequalities are satisfied
by any local hidden variable theory, but they may be violated by quantum
mechanics. However, Shunlong Luo and Shuangshuang Fu have introduced a new way
to quantify nonlocality by measurement, which is called the MIN \cite{Luo}, and following their
treatise, a number of papers emerged to perfect its definition
\cite{Xi,Sayyed} and discussed its properties \cite{Sen,Guo}. In
addition, some authors have analyzed its dynamical behavior and
compared it with other quantum correlation measurements such as the
geometric discord \cite{Hu,Zhang}. However, all of these studies
don't involve the effect on the MIN resulting from relativistic
effect. In fact, the study that how the Unruh effect \cite{Unruh}
affects the MIN can help us
implement the quantum task better and more efficient. Inspired by
that, in this paper we analyze how the Unruh effect affects the
MIN for both the Dirac and Bosonic fields and find some new properties.

Our paper is constructed as follows. In section \ref{definition of MIN} we simply introduce the definition of the MIN.
In sections \ref{Dirac fields MIN} and \ref{Bosonic fields MIN} how the Unruh effect affects the MIN for Dirac
and Bosonic fields is respectively studied. And we summarize and discuss our conclusions in the last section.

\section{definition of measurement-induced-nonlocality}\label{definition of MIN}

Recently, Luo $\emph{et al}$ \cite{Luo} have introduced a way  to
quantify nonlocality from a geometric perspective in terms of
measurements, which is called the MIN. For a bipartite quantum state
$\rho$ shared by subsystem $A$ and $B$ with respective system
Hilbert space $H^A$ and $H^B$, we can find the difference between
the overall pre- and post-measurement states by
performing a local von Neumann measurements on part $A$. To capture
the genuine nonlocal effect of measurements on the state, the key
point is that the measurements do not disturb the local state
$\rho^A=tr_B\rho$. Based on this idea, the MIN can be defined by
\begin{eqnarray}\label{intial MIN}
N(\rho)=\max_{\Pi^A}\parallel\rho-\Pi^A(\rho)\parallel^2.
\end{eqnarray}
For a general $2\times2$ dimensional system
\begin{eqnarray}\label{Bloch representation}
\rho=\frac{1}{2}\frac{\mathbf{1}^A}{\sqrt{2}}\otimes\frac{\mathbf{1}^B}{\sqrt{2}}
+\sum^3_{i=1}x_iX_i\otimes\frac{\mathbf{1}^B}{\sqrt{2}}
+\frac{\mathbf{1}^A}{\sqrt{2}}\otimes\sum^3_{j=1}y_jY_j
+\sum^3_{i=1}\sum^3_{j=1}t_{ij}X_i\otimes Y_j,
\end{eqnarray}
its MIN is given by \cite{Luo}
\begin{eqnarray}\label{MIN}
N(\rho)=
\left\{
\begin{array}{lr}
trTT^t-\frac{1}{\|\mathbf{x}\|^2}\mathbf{x}^tTT^t\mathbf{x}  & ~~{\text{if}}~~ \mathbf{x}\neq0,\\
trTT^t-\lambda_3  & ~~{\text{if}}~~ \mathbf{x}=0,
\end{array}
\right.
\end{eqnarray}
where $TT^t(T=(t_{ij}))$ is a $3\times3$ dimensional matrix,
$\lambda_3$ is its minimum eigenvalue, and
$\|\mathbf{x}\|^2=\sum_ix^2_i$ with $\mathbf{x}=(x_1,x_2,x_3)^t$.

\section{MIN for Dirac fields}\label{Dirac fields MIN}

Two sets of modes named Rindler and Minkowski modes are corresponding to the solutions of Dirac equation in Rindler and
Minkowski space-time, respectively. It is found that the combinations of Minkowski modes called Unruh modes can be transformed into monochromatic Rindler modes. Besides, they share the same vacuum state as the standard monochromatic Minkowski modes. Therefore, the
annihilation operators of Unruh modes can also annihilate the Minkowski vacuum, and they are give by
\begin{eqnarray}\label{annihilation operators}
C_{\Omega,R/L}=\frac{1}{(e^{\Omega/T}+1)^{1/2}}(e^{\Omega/2T}c_{\Omega,I/II}-d^\dagger_{\Omega,II/I}),
\end{eqnarray}
where $c^\sigma_{\Omega,\varsigma}$ and $d^\sigma_{\Omega,\varsigma}$ with $\sigma=\{\dagger,-\}$ and $\varsigma=\{I,II\}$ denote
that particle and antiparticle operators in region $\varsigma$, respectively, and $T$ is Unruh temperature.

Going beyond the single mode approximation \cite{Luoko}, we consider a general Unruh mode being an arbitrary combination of
the independent Unruh modes (\ref{annihilation operators}) such as
\begin{eqnarray}\label{Unruh operators}
c^\dagger_{\Omega,U}=q_L(C^\dagger_{\Omega,L}\otimes\mathbf{1}_R)+q_R(\mathbf{1}_L\otimes C^\dagger_{\Omega,R}).
\end{eqnarray}
By using the creation operator to act on the vacuum, the associated single particle state is easily obtained.

It is needed to note that the Unruh modes are spatial delocalized, so can not be completely measured. Additionally near the acceleration horizon they are highly oscillatory from the inertial observer's viewpoint. This implies that they are physically unfeasible states. However, using the Unruh modes, we can get a family of solutions only depending on the parameterized acceleration $r$ (or the Unruh temperature $T$). The entangled state constructed by them will be the maximally entangled in the inertial limit, and from the perspective of accelerated observers it is single-frequency entangled state. Moreover, as argued in Ref. \cite{Brown}, the use of Unruh modes to extract conclusions regarding fundamental physics can be justified according to some desirable features, such as they are purely positive frequency linear combinations of Minkowski modes, and share the same vacuum with Minkowski monochromatic modes, and so on. These features ensure that the parameterized acceleration dependent states can describe the behavior of maximally entangled states in small acceleration limit, and they also yield the correct behavior for physical states in the large acceleration limit, so a main result (different behaviors between the Bosonic fields and Dirac fields in the infinite accelerated limit) in our paper can also applies to the physical states. Furthermore, such results qualitatively describe the behavior for the relevant states as a function of the acceleration. For these reasons, the Unruh modes are commonly used in literatures.

Usually, the Unruh monochromatic mode $|0_\Omega\rangle_\mathrm{U}$, from the non-inertial observers' perspective, can be expressed as \cite{Montero,Montero1}
\begin{eqnarray}\label{vacuum}
|0_\Omega\rangle_\mathrm{U}=\frac{1}{e^{\Omega/T}+1}\left(e^{\Omega/T}|0000\rangle_\Omega-e^{\Omega/2T}|0011\rangle_\Omega
+e^{\Omega/2T}|1100\rangle_\Omega-|1111\rangle_\Omega\right),
\end{eqnarray}
where the notation $|p_\Omega\rangle^\dagger_I|q_\Omega\rangle^-_{II}|m_\Omega\rangle^-_I|q_\Omega\rangle^\dagger_{II}$ is
used. Likewise, the particle state of Unruh mode $\Omega$ in the Rindler basis is found to be
\begin{eqnarray}\label{particle state}
\nonumber
|1_\Omega\rangle_\mathrm{U}=\frac{q_R}{(e^{\Omega/T}+1)^{1/2}}\left(e^{\Omega/2T}|1000\rangle_\Omega-|1011\rangle_\Omega\right)
\\
+\frac{q_L}{(e^{\Omega/T}+1)^{1/2}}\left(e^{\Omega/2T}|0001\rangle_\Omega+|1101\rangle_\Omega\right).
\end{eqnarray}

Noting that different ordering of operators leads to Fermionic entanglement ambiguity in non-inertial frames because
of the anticommutation relation satisfied by Fermionic operators. However, only one of these ordering gives physical
results, as done in ref. \cite{Montero1}, we adopt the physical ordering in our paper.

\subsection{MIN shared by Alice and Rob}

We now assume that Alice and Rob share a X-type initial state
\begin{equation}\label{initial states}
\rho_{AB}=\frac{1}{4}\left(I_{AB}+ \sum_{i=1}^{3}c_{i}\sigma_{i}%
^{(A)}\otimes\sigma_{i}^{(B)}\right),
\end{equation}
where $I_{A(B)}$ is the identity operator in subspace $A(B)$,  and
$\sigma_{i}^{(n)}$ is the Pauli operator in direction $i$ acting on
the subspace  $n=A,B$, $c_{i} \in\mathfrak{R}$ such that $0\leq \mid
c_{i}\mid\leq1$ for $i=1,2,3$. Obviously, Eq.  (\ref{initial states})
represents a class of states including the well-known initial
states, such as the Werner initial state ($\left\vert
c_{1}\right\vert =\left\vert c_{2}\right\vert =\left\vert
c_{3}\right\vert =\alpha$), and Bell basis state ($\left\vert
c_{1}\right\vert =\left\vert c_{2}\right\vert =\left\vert
c_{3}\right\vert =1$).

After the coincidence of Alice and Rob, Alice stays stationary
while Rob moves with an uniform acceleration $a$. To describe the
states shared by these two relatively accelerated observers in
detail, we must use Eqs.  (\ref{vacuum}) and (\ref{particle state}) to
rewrite Eq.  (\ref{initial states}) in terms of Minkowski modes for
Alice, Rindler modes I for Rob and Rindler modes II for Anti-Rob,
which implies that Rob and Anti-Rob are respectively confined in
region $\mathrm{I}$ and $\mathrm{II}$. The regions $\mathrm{I}$ and
$\mathrm{II}$ are causally disconnected, and the information which
is physically accessible to the observers is encoded in the
Minkowski modes $A$ and Rindler modes $\mathrm{I}$, but the
physically unaccessible information is encoded in the Minkowski
modes $A$ and Rindler modes $\mathrm{II}$. So we must trace over the
Rindler modes $\mathrm{II}$ (modes $\mathrm{I}$) when we only
consider the physically accessible (unaccessible) information.

We first consider the MIN between modes $A$ and particle modes in region $\mathrm{I}$.  By
taking the trace over the states of region $\mathrm{II}$ and antiparticle states of region $\mathrm{I}$, we obtain
the reduced density matrix of Alice-Rob modes
\begin{eqnarray} \label{state of Alice and Rob}
\nonumber\rho_{A,I}=\frac{1}{4}
\left(
  \begin{array}{cccc}
    \frac{1+q^2_L+c_3q^2_R}{e^{-\Omega/T}+1} & 0 & 0 &  \frac{(c_1-c_2)q_R}{(e^{-\Omega/T}+1)^{\frac{1}{2}}} \\
    0 & (1-c_3)q^2_R+\frac{1+c_3q^2_R+q^2_L}{e^{\Omega/T}+1} &  \frac{(c_1+c_2)q_R}{(e^{-\Omega/T}+1)^{\frac{1}{2}}}   & 0 \\
    0 &  \frac{(c_1+c_2)q_R}{(e^{-\Omega/T}+1)^{\frac{1}{2}}}  & \frac{1-c_3q^2_R+q^2_L}{e^{-\Omega/T}+1}  & 0 \\
    \frac{(c_1-c_2)q_R}{(e^{-\Omega/T}+1)^{\frac{1}{2}}}  & 0 & 0 &  (1+c_3)q^2_R+\frac{1-c_3q^2_R+q^2_L}{e^{\Omega/T}+1}\\
  \end{array}
\right)
,
\end{eqnarray}
where $|mn\rangle=|m\rangle_{\mathrm{A}}|n_\Omega\rangle^\dagger_{\mathrm{I}}$.
For convenience to calculate the MIN, we rewrite the state
$\rho_{A,\mathrm{I}}$ in terms of Bloch representation, which is
given by
\begin{eqnarray}\label{AI Bloch representation}
\rho_{A,I}=\frac{1}{4}\left( \mathbf{1}_A\otimes\mathbf{1}_I
+c'_0\mathbf{1}_A\otimes\sigma^{(I)}_3
+\sum^3_{i=1}c'_i\sigma^{(A)}_i\otimes\sigma^{(I)}_i \right),
\end{eqnarray}
where $c'_0=\frac{q^2_Le^{\Omega/T}-1}{(e^{\Omega/T}+1)}$, $c'_1=\frac{c_1q_R}{(e^{-\Omega/T}
+1)^{\frac{1}{2}}}$, $c'_2=\frac{c_2q_R}{(e^{-\Omega/T}+1)^{\frac{1}{2}}}$
and $c'_3=\frac{c_3q^2_R}{(e^{-\Omega/T}+1)}$. From Eq. (\ref{MIN}), it is easy
to get the MIN for the state $\rho_{A,I}$
\begin{eqnarray}\label{MIN for AI}
N(\rho_{A,I})=\frac{1}{4}\{\frac{(c_1q_R)^2}{(e^{-\Omega/T}+1)}
+\frac{(c_2q_R)^2}{(e^{-\Omega/T}+1)}+\frac{(c_3q^2_R)^2}{(e^{-\Omega/T}+1)^2}
\nonumber \\
-\min[\frac{(c_1q_R)^2}{(e^{-\Omega/T}+1)}
,\frac{(c_2q_R)^2}{(e^{-\Omega/T}+1)},\frac{(c_3q^2_R)^2}{(e^{-\Omega/T}+1)^2}]\}.
\end{eqnarray}
Obviously, $\min[\frac{(c_1q_R)^2}{(e^{-\Omega/T}+1)}
,\frac{(c_2q_R)^2}{(e^{-\Omega/T}+1)},\frac{(c_3q^2_R)^2}{(e^{-\Omega/T}+1)^2}]$ depends on both the
coefficients $c_i$ of the states in Eq. (\ref{initial states}) and
the Unruh temperature. Besides, according to the definition of geometric discord
in Ref. \cite{Bor,Adesso}, we can also get the geometric discord from Eq.  (\ref{MIN for AI}) only
by changing $\min$ with $\max$, so it is no doubt that the MIN is always greater than
or equal to the geometric discord. And it should be note that the geometric discord with $c_i=1$ and
$q_R=1$ will go back to the case in Ref. \cite{Brown} which discussed the geometric
discord by using the single-mode approximation.

For X-type initial state, the MIN gives different dynamic behaviors for different $c_i$:

(i) If $|c_1|,|c_2|\geq|c_3q^2_R|$ in Eq. (\ref{initial states}), the
minimum term in Eq. (\ref{MIN for AI}) is
$\frac{(c_3q^2_R)^2}{(e^{-\Omega/T}+1)^2}$. In this case, the MIN, for fixed $c_3$ and $q_R$, decreases monotonously as the Unruh temperature
increases.

(ii) For the case of $|c_3q^2_R|>\min\{|c_1|,|c_2|\}$ and both $c_1$  and
$c_2$ don't equals to 0 at the same time, if
$\min\{|c_1|,|c_2|\}\geq\frac{\sqrt{2}}{2}|c_3 q^2_R|$, the MIN has a
peculiar dynamics with a sudden change as the Unruh temperature
increases, i.e., $N(\rho_{A,I})$ decays quickly until
\begin{eqnarray}\label{TSC}
 T_{sc}=\frac{-\Omega}{\ln(\frac{|c_3 q_R|^2}{min\{|c_1|^2,|c_2|^2\}}-1)},
\end{eqnarray}
and then $N(\rho_{A,I})$ decays relatively slowly. Otherwise,  the
MIN decays monotonously as the temperature increases.

(iii) Finally, if $|c_1|=|c_2|=0$, we have a monotonic decay  of
$N(\rho_{A,I})$.

The decrease of the MIN means that the difference between the pre-
and post-measurement states becomes smaller, i.e., the disturbance
induced by local measurement weaken. If we understand the MIN as
some kind of correlations, this decrease means that the quantum
correlation shared by two relatively accelerated observers decreases,
i.e., less quantum resource can be used for the quantum information task
by these two observers. So the Unruh effect affects quantum communication
process by inducing the decrease of quantum resource.

\begin{figure}[htp!]
\centering
\includegraphics[width=0.5\textwidth]{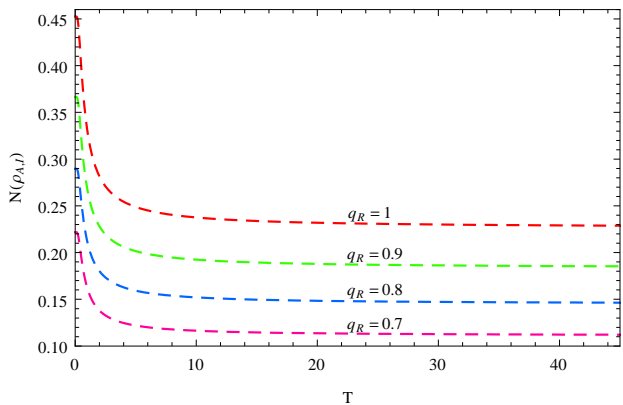}
\caption{(Color online) The MIN of state $\rho_{A,I}$  as a function
of Unruh temperature $T$ for different choices of $q_R$.
We take parameters $c_1=1$, $c_2=0.9$ and
$|c_3q^2_R|\leq|c_1|,|c_2|$ here.}\label{MINA1}
\end{figure}

\begin{figure}[htp!]
\centering
\includegraphics[width=0.8\textwidth]{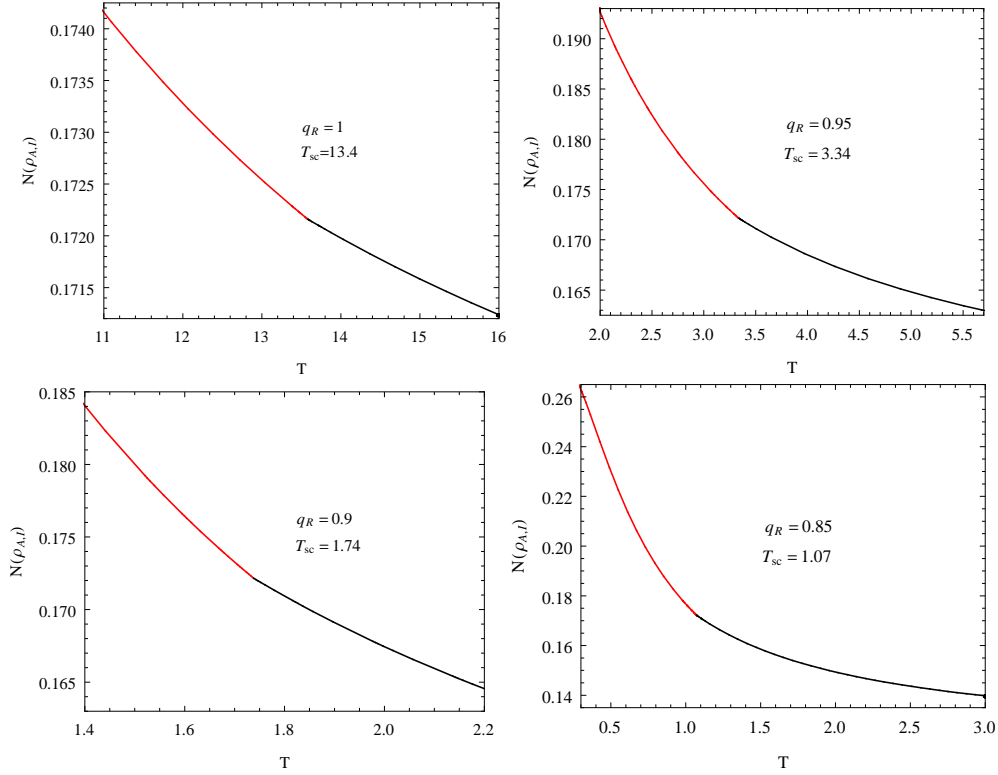}
\caption{(Color online) The details of sudden change of MIN,
and the joint point between red line and black line is the sudden
change point. Here we take $c_1 =0.9$, $c_2=0.72$, and
$c_3=1$.}\label{MINA2}
\end{figure}

By taking $\Omega=1$ hereafter, we show the dynamical behavior of
$N(\rho_{A,I})$ in Fig. \ref{MINA1} and \ref{MINA2}. Fig. \ref{MINA1}
exhibits one example of case (i) for different $q_R$, obviously, smaller
$q_R$ corresponds to less MIN, but it is no fundamental difference in the
degradation of MIN for different choices of $q_R$. Fig \ref{MINA2} tells us some details of
the sudden change occurring for case (ii), and the smaller $q_R$ is, the  earlier
the sudden change occurs. Furthermore,
We find that the MIN,
as the Unruh temperature approaches to the infinite,
has a limit
\begin{eqnarray}\label{limit of N}
\lim_{T\rightarrow\infty}N(\rho_{A,I})= \frac{1}{16}\{2(c_1q_R)^2
+2(c_2q_R)^2+(c_3q^2_R)^2-\min[2(c_1q_R)^2,2(c_2q_R)^2,(c_3q^2_R)^2]\}.
\end{eqnarray}
That is to say, as long as the initial MIN does  not equal to zero,
it can persist for arbitrary Unruh temperature.

\begin{figure}[htp!]
\centering
\includegraphics[width=0.5\textwidth]{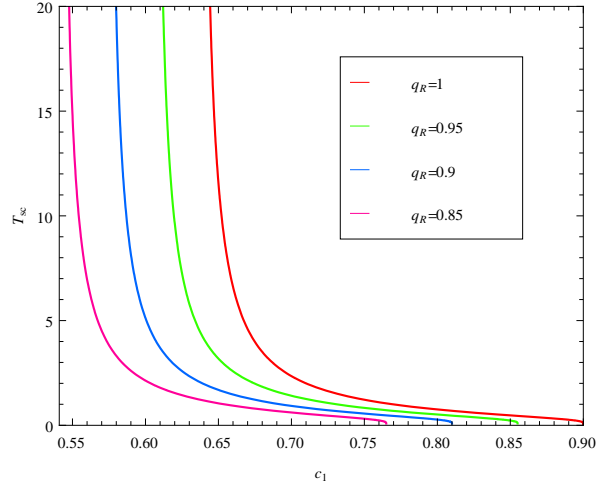}
\caption{(Color online) The $T_{sc}$ as a function of  $c_1$, here
we take $|c_1|\leq|c_2|$ and $c_3=0.9$.}\label{Tsc}
\end{figure}

We now study $T_{sc}$ of Eq.  (\ref{TSC}). If $|c_1|\leq|c_2|$,
by taking fixed $c_3$ and $q_R$, we plot how the parameter $c_1$ affects it
in Fig. \ref{Tsc}, which shows that it decreases
monotonously as $c_1$ increases. That is to say, the bigger $c_1$
is, the sudden change behavior occurs earlier. Besides, $q_R$ doesn't
lead to the fundamental difference in the degradation of $T_{sc}$
with the increase of $c_1$,
but different $q_R$ will result in different area of $c_1$ in which that sudden
change can happen. And when $|c_2|\leq|c_1|$, it is
interesting to note that with the increase of $c_2$ it
decreases monotonously too.

\subsection{MIN shared by Alice and Anti-Rob}

Then we consider the MIN between modes $\mathrm{A}$  and antiparticle modes in region
$\mathrm{II}$. By tracing over all modes in region $\mathrm{I}$ and particle modes in region
$\mathrm{II}$, we get the reduced density matrix of Alice-antiRob modes
\begin{eqnarray}\label{AII Bloch representation}
\rho_{A,II}=\frac{1}{4}\left( \mathbf{1}_A\otimes\mathbf{1}_{II}
+c'_0\mathbf{1}_A\otimes\sigma^{(II)}_3
+\sum^3_{i=1}c'_i\sigma^{(A)}_i\otimes\sigma^{(II)}_i \right),
\end{eqnarray}
where $c'_0=\frac{e^{\Omega/T}-q^2_R}{(e^{\Omega/T}+1)}$, $c'_1=\frac{c_1q_R}{(e^{\Omega/T}
+1)^{\frac{1}{2}}}$, $c'_2=\frac{-c_2q_R}{(e^{\Omega/T}+1)^{\frac{1}{2}}}$
and $c'_3=\frac{-c_3q^2_R}{(e^{\Omega/T}+1)}$. Similarly, the MIN of state
$\rho_{A,II}$ can be obtained according to Eq. (\ref{MIN}), which is
\begin{eqnarray}\label{MIN for AII}
N(\rho_{A,II})=\frac{1}{4}\{\frac{(c_1q_R)^2}{(e^{\Omega/T}+1)}
+\frac{(c_2q_R)^2}{(e^{\Omega/T}+1)}+\frac{(c_3q^2_R)^2}{(e^{\Omega/T}+1)^2}
\nonumber \\
-\min[\frac{(c_1q_R)^2}{(e^{\Omega/T}+1)},\frac{(c_2q_R)^2}{(e^{\Omega/T}+1)},
\frac{(c_3q^2_R)^2}{(e^{\Omega/T}+1)^2}]\}.
\end{eqnarray}
We can also obtain the geometric discord of Alice-antiRob modes by changing
$\min$ with $\max$ in Eq.  (\ref{MIN for AII}). Apparently, it is always smaller
than or equal to the MIN.

By analysing Eq.  (\ref{MIN for AII}), we will find that:

(i) If $|c_1|,|c_2|\geq|c_3q^2_R|$, the MIN increases monotonously as the
Unruh temperature increases for fixed $c_3$ and $q_R$.

(ii) For the case of $|c_3q^2_R|>\min\{|c_1|,|c_2|\}$ and both $c_1$ and
$c_2$ don't equal to 0 at the same time, if
$\min\{|c_1|,|c_2|\}\leq\frac{\sqrt{2}}{2}|c_3q^2_R|$, the MIN has a
peculiar dynamics with a sudden change at $T_{sc}$
\begin{eqnarray}\label{TSC2}
T_{sc}=\frac{\Omega}{\ln(\frac{|c_3q_R|^2}{min\{|c_1|^2,|c_2|^2\}}-1)}.
\end{eqnarray}
Otherwise, the MIN increases monotonously with the increase of the
Unruh temperature.

(iii) Finally, if $|c_1|=|c_2|=0$, we have a monotonic increase  of
$N(\rho_{A,II})$.

\begin{figure}[htp!]
\centering
\includegraphics[width=0.5\textwidth]{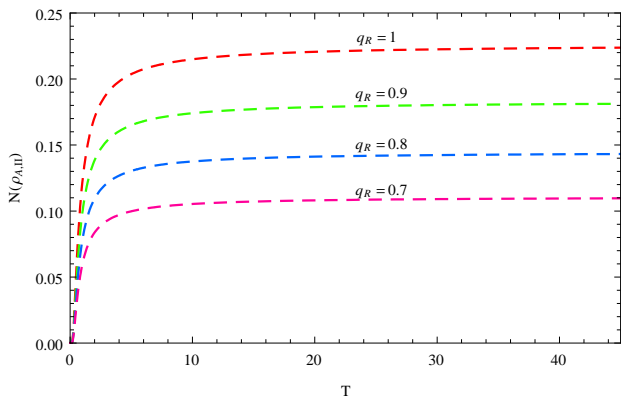}
\caption{(Color online) The MIN of state $\rho_{A,II}$  as a
function of Unruh temperature $T$.  We take parameters $c_1=1$,
$c_2=0.9$ and $|c_3q^2_R|\leq|c_1|,|c_2|$ with different $q_R$.}\label{MINII1}
\end{figure}

\begin{figure}[htp!]
\centering
\includegraphics[width=0.8\textwidth]{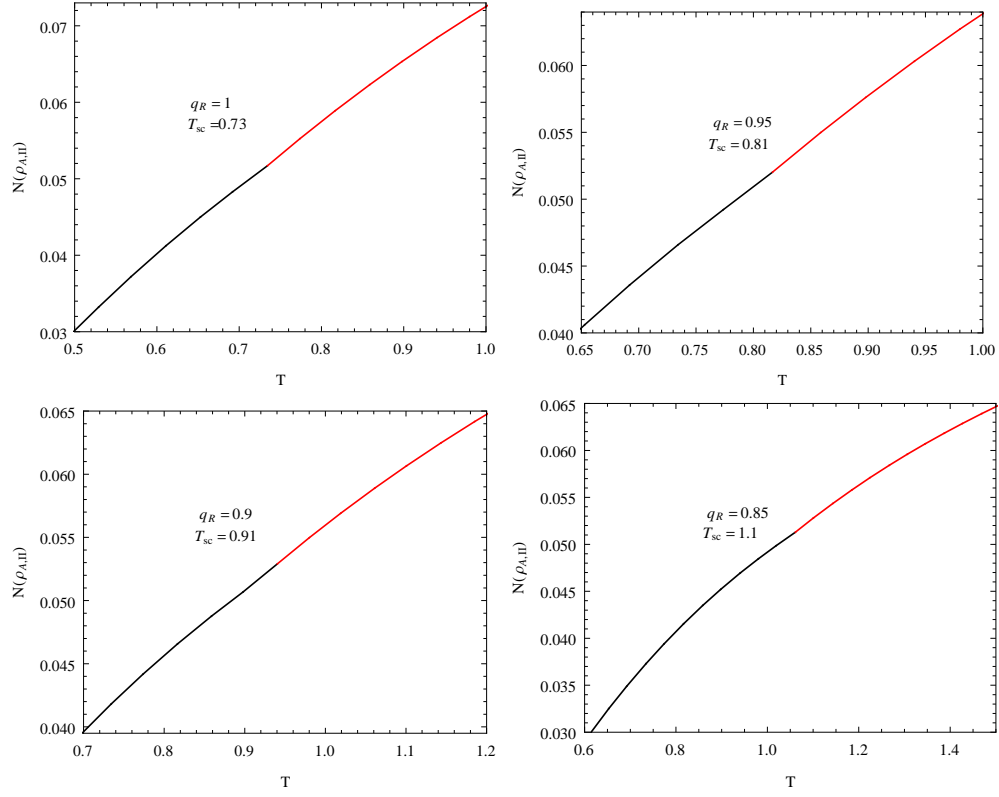}
\caption{(Color online) The details of sudden change of MIN for
state $\rho_{A,II}$, and the joint point between red line and
black line is the sudden change point. Here we take $c_1 =0.9$,
$c_2=0.45$, and $c_3=1$.}\label{MINII2}
\end{figure}

In Fig.  \ref{MINII1} and \ref{MINII2} we plot $N(\rho_{A,II})$
versus the Unruh temperature. And Fig. \ref{MINII1} shows one example of case (i),
we find that the MIN, as the Unruh temperature increases, increases monotonously,
and as the Unruh temperature approaches to the infinite limit, it is close to
\begin{eqnarray}
\lim_{T\rightarrow\infty}N(\rho_{A,II})=\frac{1}{16}\{2(c_1q_R)^2
+2(c_2q_R)^2+(c_3q^2_R)^2-\min[2(c_1q_R)^2,2(c_2q_R)^2,(c_3q^2_R)^2]\},
\end{eqnarray}
which is the same as $\lim_{T\rightarrow\infty}N(\rho_{A,I})$. In
addition, as $T=0$ the MIN vanishes, which means that the
correlation of Alice-antiRob modes is local when the observers are inertial. Obviously, bigger
$q_R$ corresponds to bigger MIN. And Fig.  \ref{MINII2} tells us that the sudden
change, for fixed $c_i$, comes later for smaller $q_R$.

\begin{figure}[htp!]
\centering
\includegraphics[width=0.5\textwidth]{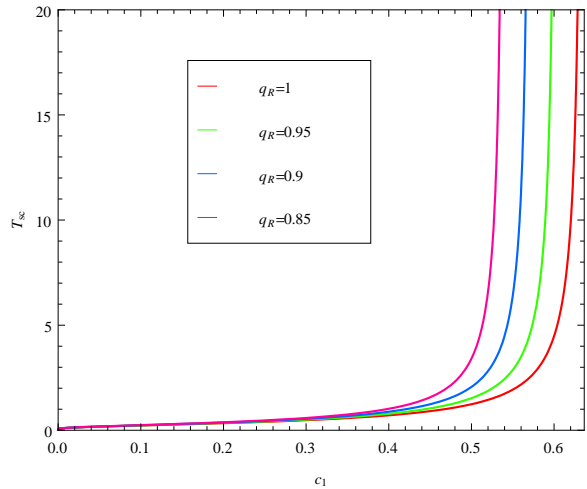}
\caption{(Color online) The $T_{sc}$ as a function  of $c_1$, here
we take $|c_1|\leq|c_2|$ and $c_3=0.9$.}\label{Tsc1}
\end{figure}

We study $T_{sc}$ of Eq. (\ref{TSC2}), when $|c_1|<|c_2|$, for fixed $c_3$ and $q_R$, we plot $T_{sc}$ as a function of $c_1$
in Fig.  \ref{Tsc1}. We learn from the figure that, unlike the Fig. \ref{Tsc},
 $T_{sc}$ increases monotonously with the increase of $c_1$. That is to say,
the bigger $c_1$ is, the sudden change behavior occurs latter. Furthermore,
$q_R$ can change the area of $c_1$ in which that sudden change occurs.
And when $|c_2|<|c_1|$, it is also important to note that as $|c_2|$ increases
$T_{sc}$ increases monotonously too.

\subsection{Relating MIN to Nonlocality}

Because the MIN is introduced to describe non-locality, for further
understanding it we will compare it with the maximal possible value
$\langle B_{max}\rangle$ of the Bell-CHSH expectation value.

As shown in Ref. \cite{Nicolai}, the $\langle B_{max}\rangle$  for a
given state $\rho$ is determined by
\begin{eqnarray}\label{Bell expectation value}
\langle B_{max}\rangle_\rho=2\sqrt{\mu_1+\mu_2},
\end{eqnarray}
where $\mu_1$, $\mu_2$ are the two largest eigenvalues  of
$U(\rho)=TT^t$, the matrix $T=(t_{ij})$ with $t_{ij}=Tr[\rho\sigma_i\otimes\sigma_j]$.

Using Eqs. (\ref{AI Bloch representation}) and (\ref{Bell expectation
value}), $\langle
B_{max}\rangle_{\rho_{A,I}}$ is given by
\begin{eqnarray}\label{Bmax}
\langle B_{max}\rangle_{\rho_{A,I}}=2\{\frac{(c_1q_R)^2}{(e^{-\Omega/T}+1)}
+\frac{(c_2q_R)^2}{(e^{-\Omega/T}+1)}+\frac{(c_3q^2_R)^2}{(e^{-\Omega/T}+1)^2}
\nonumber \\
-\min[\frac{(c_1q_R)^2}{(e^{-\Omega/T}+1)},\frac{(c_2q_R)^2}{(e^{-\Omega/T}+1)},
\frac{(c_3q^2_R)^2}{(e^{-\Omega/T}+1)^2}]\}^{1/2},
\end{eqnarray}

\begin{figure}[htp!]
\centering
\includegraphics[width=0.5\textwidth]{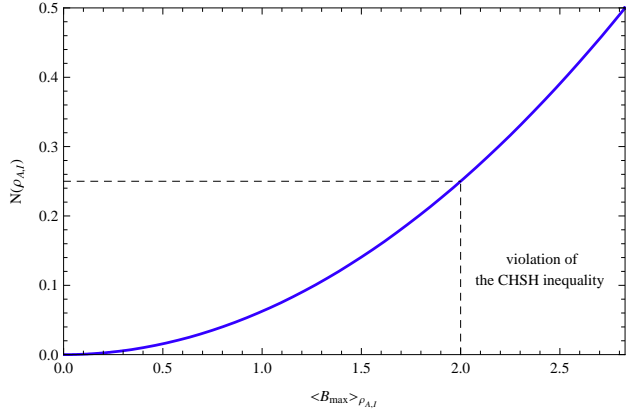}
\caption{(Color online) The MIN of state $\rho_{A,I}$  as function
of the maximally possible value of the Bell-CHSH expectation
value.}\label{NB relation}
\end{figure}

It is interesting to note that
\begin{eqnarray}
N(\rho_{A,I})=\frac{1}{16}\langle B_{max}\rangle_{\rho_{A,I}}^2.
\end{eqnarray}
We plot $N(\rho_{A,I})$ versus $\langle B_{max}\rangle_{\rho_{A,I}}$
in Fig. \ref{NB relation}, which shows that $N(\rho_{A,I})$ increases
monotonously as $\langle B_{max}\rangle_{\rho_{A,I}}$ increases and
it vanishes at zero point of $\langle B_{max}\rangle_{\rho_{A,I}}$.
It is well known that Bell inequality must be obeyed by local realism
theory, but may be violated by quantum mechanics. If we get
$\langle B_{max}\rangle_{\rho_{A,I}}>2$, it means that the
violation of Bell-CHSH inequality, which tells us that there exists nonlocal
quantum correlation. But when $\langle B_{max}\rangle_{\rho_{A,I}}\leq2$, it doesn't
mean that no quantum correlation exists, at least for some mixed states, which have quantum
correlation but obey the Bell inequality. So we can't be sure that whether quantum
correlation exists or not when $\langle B_{max}\rangle_{\rho_{A,I}}\leq2$.
However, the MIN, which is an indicator of the global effect caused by locally
invariant measurement, is introduced to quantify nonlocality, nonzero MIN
means existence of nonlocality. And from Fig. \ref{NB relation}
we see that the MIN persists for all $\langle B_{max}\rangle_{\rho_{A,I}}$
except for zero. Thus, the MIN, understood as some kind of correlations, is more general
than the quantum nonlocality related to violation of the Bell's inequalities.

\section{MIN for Bosonic fields}\label{Bosonic fields MIN}

For Bosonic field, we can also get the Unruh vacuum in terms of Rindler vacuum and
its excitations. As shown in Ref. \cite{Luoko}, the monochromatic Unruh vacuum mode can
be expressed as
\begin{eqnarray}\label{Bosonic expansion}
|0_\Omega\rangle_U=\frac{1}{\cosh r_\Omega}\sum^\infty_{n=0} \tanh^nr_\Omega
|n_\Omega\rangle_\mathrm{I}|n_\Omega\rangle_{\mathrm{II}},
\end{eqnarray}
with $\tanh r_\Omega=e^{-\frac{\Omega}{2T}}$. And a positive frequency particle state
is given by
\begin{eqnarray}\label{Bosonic particle state}
\nonumber
A^\dagger_{\Omega,\mathrm{U}}|0_\Omega\rangle_\mathrm{U}&=&|1_\Omega\rangle_\mathrm{U}
\\
&=&\sum^\infty_{n=0}\frac{\tanh^nr_\Omega}{\cosh r_\Omega}
\frac{\sqrt{n+1}}{\cosh r_\Omega}
(q_R|(n+1)_\Omega\rangle_\mathrm{I}|n_\Omega\rangle_\mathrm{II}+
q_L|n_\Omega\rangle_\mathrm{I}|(n+1)_\Omega\rangle_\mathrm{II}).
\end{eqnarray}

For simplicity, we choose the Bell state
\begin{eqnarray}\label{Bell state}
|\Psi\rangle=\frac{1}{\sqrt{2}}(|0_\omega\rangle_\mathrm{M}|0_\Omega\rangle_\mathrm{U}
+|1_\omega\rangle_\mathrm{M}|1_\Omega\rangle_\mathrm{U})
\end{eqnarray}
as initial state and study the MIN of both Alice-Rob modes and Alice-antiRob modes. After rewriting Rob's modes in
Eq.  (\ref{Bell state}) in terms of the Rindler basis, the Alice-Rob density matrix
is obtained by tracing over region $\mathrm{II}$, with the result,
\begin{eqnarray}\label{Bosonic final state}
\rho_{A,I}=\frac{1}{2}\sum^\infty_{n=0}\left[\frac{\tanh^nr_\Omega}{\cosh r_\Omega}\right]^2\rho^n_{A,R},
\end{eqnarray}
where
\begin{eqnarray}
\nonumber
\rho^n_{A,R}=&&|0n\rangle\langle0n|+\frac{n+1}{\cosh^2r_\Omega}(|q_R|^2|1n+1\rangle\langle1n
+1|+|q_L|^2|1n\rangle\langle1n|)
\\ \nonumber
&&+\frac{\sqrt{n+1}}{\cosh r_\Omega}(q_R|1n+1\rangle\langle0n|+q_L\tanh r_\Omega|1n\rangle\langle0n+1|)
+\frac{\sqrt{(n+1)(n+2)}}{\cosh^2r_\Omega}
\\ \nonumber
&&\times q_Rq^\ast_L\tanh r_\Omega|1n+2\rangle\langle1n|+(\mathrm{H.c.})_{\mathrm{nondiag}},
\end{eqnarray}
where $(\mathrm{H.c.})_{\mathrm{nondiag}}$ means Hermitian conjugate of only the non-diagonal
terms. And it is necessary to note that the Alice-antiRob density matrix, due to the symmetry
in the Unruh modes between regions $\mathrm{I}$ and $\mathrm{II}$, can be gained by exchanging
$q_R$ and $q_L$.

It is convenient to rewrite state (\ref{Bosonic final state}) in the following form,
\begin{eqnarray}
\rho_{A,I}=\frac{1-t^2}{2}(|0\rangle\langle0|\otimes M_{00}+|1\rangle\langle1|\otimes M_{11}
+|0\rangle\langle1|\otimes M_{01}+|1\rangle\langle0|\otimes M_{10}),
\end{eqnarray}
where $t=\tanh r_\Omega$ and the matrices on Rob's Hilbert space are
\begin{eqnarray}\label{M}
\nonumber
M_{00}&=&\sum^\infty_{n=0}t^{2n}|n\rangle\langle n|,
\\ \nonumber
M_{11}&=&(1-t^2)\sum^\infty_{n=0}t^{2n}[(n+1)(|q_R|^2|n+1\rangle\langle n+1|+|q_L|^2|n\rangle\langle n|)
\\ \nonumber
&&+t\sqrt{(n+1)(n+2)}(q_Rq^\ast_L|n+2\rangle\langle n|+q^\ast_Rq_L|n\rangle\langle n+2|)],
\\ \nonumber
M_{01}&=&\sqrt{1-t^2}\sum^\infty_{n=0}\sqrt{n+1}t^{2n}(q^\ast_R|n\rangle\langle n+1|+q^\ast_L|n+1\rangle\langle n|),\\ \nonumber
M_{10}&=&M^\dagger_{01}.
\end{eqnarray}

We make a projective measurement on the qubit of Alice, and after the measurement the final state is given by
\begin{eqnarray}
\rho'_{A,I}=\sum_{\alpha=\pm}(\Pi_\alpha\otimes \mathbf{1}_I)\rho_{A,I}(\Pi_\alpha\otimes \mathbf{1}_I)=
\sum_{\alpha=\pm}p_\alpha\Pi_\alpha\otimes\rho_{R|\alpha}.
\end{eqnarray}
Here, the projectors
\[\Pi_{\pm}=\frac{1}{2}[(1\pm x_3)|0\rangle\langle0|+(1\mp x_3)|1\rangle\langle1|
\pm(x_1-ix_2)|1\rangle\langle0|\pm(x_1+ix_2)|0\rangle\langle1|],\]
$\rho_{R|\alpha}$ denotes the post-measured state of Rob's reduced system conditioned
on the outcome $\alpha$, and $p_\alpha$ is the corresponding probability.

After a series of calculations, we finally get the difference between the pre- and post-
measurement states, with the result,
\begin{eqnarray}\label{Bosonic MIN}
\nonumber
\mathrm{Tr}((\rho_{A,I}-\rho'_{A,I})^2)&=&\frac{(1-t^2)^2}{8}[(1-x^2_3)(\mathrm{Tr}(M^2_{00})
+\mathrm{Tr}((M^2_{11}))-2\mathrm{Tr}(M_{00}M_{11}))
\\ \nonumber
&&+2(1+x^2_3)\mathrm{Tr}(M_{01}M_{10})]
\\ \nonumber
&=&\frac{(1-t^2)^2}{8}[\frac{1+x^2_3}{1+t^2}+\frac{2(1-t^2)(|q_R|^2-|q_L|^2)}{1+t^2}
\\
&&+\frac{1-x^2_3}{(1+t^2)^3}((1+t^4)-2(1-t^2)^2|q_R|^2|q_L|^2)].
\end{eqnarray}
Therefore, according to the definitions of MIN \cite{Luo} and geometric discord \cite{Bor,Adesso},
Eq.  (\ref{Bosonic MIN}) with $x_3=1$ and $x_3=0$ gives the MIN and geometric discord, respectively.
When $x_3=0$ and $q_R=1$, our result can go back to the case in Ref. \cite{Brown} where the geometric discord
is obtained by using the single-mode approximation. Furthermore, we can also obtain the MIN and
geometric discord of Alice-antiRob modes by exchanging $q_R$ and $q_L$ in Eq.  (\ref{Bosonic MIN}).
And it is interesting to note that both the MIN and geometric discord approach to zero in the infinity limit
of Unruh temperature $(t\rightarrow1)$, which is different from that of Dirac field case discussed above.

\begin{figure}[htp!]
\centering
\includegraphics[width=0.8\textwidth]{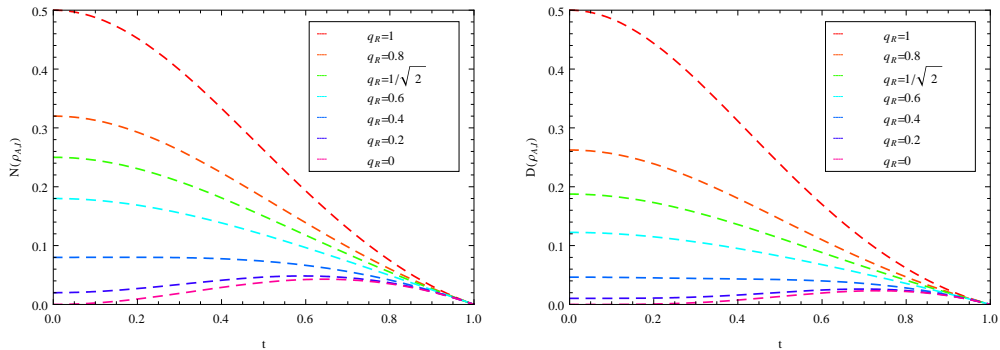}
\caption{(color online) The MIN (left one) and geometric discord (right one) as a function of
parameterized Unruh temperature $t=e^{-\frac{\Omega}{2T}}$ with fixed $q_R$.}\label{Bosonic MIN and Discord1}
\end{figure}

In Fig.  \ref{Bosonic MIN and Discord1}, we plot the MIN and geometric discord
as a function of parameterized Unruh temperature for Alice-Rob modes.
It is shown that: (i) when $0\leq q_R< q_{sc}=\frac{1}{\sqrt{6}}$ the
MIN, as $t$ increases, firstly increases to an maximal value, then decays monotonously, and finally approaches
to zero. Interestingly, when $0\leq q_R< q_{sc}\approx0.37566$ the geometric discord has the same dynamic behavior too;
and (ii) both the MIN and geometric discord, when $q_R\geq q_{sc}$, degrades monotonously with the increase of $t$.
Obviously, the dynamic behavior (i) is distinctly different from Dirac fields case that the MIN always decays.

\begin{figure}[htp!]
\centering
\includegraphics[width=0.8\textwidth]{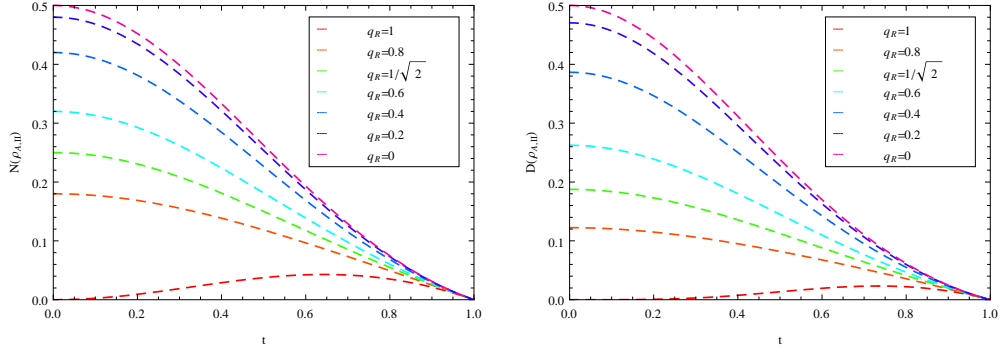}
\caption{(color online) The MIN (left one) and geometric discord (right one) as a function of
parameterized Unruh temperature $t=e^{-\frac{\Omega}{2T}}$ with fixed $q_R$.}\label{Bosonic MIN and Discord2}
\end{figure}

In Fig.  \ref{Bosonic MIN and Discord2}, we plot the MIN and geometric discord
as a function of parameterized Unruh temperature for Alice-antiRob modes.
It is shown that: (i) when $0\leq q_R\leq q_{sc}=\sqrt{\frac{5}{6}}$ the
MIN degrades monotonously with the increase of $t$, and approaches to zero
at the infinite acceleration limit. Interestingly, when $0\leq q_R\leq q_{sc}\approx0.92676$ the geometric discord has the same dynamic behavior too; (ii) both the MIN
and geometric discord, when $q_R> q_{sc}$, firstly increases to an maximal
value, then decays monotonously, and finally approaches to zero. And it is
interesting to note that these dynamic behaviors are in sharp contrast to
the Dirac fields case that the MIN always increases.

The obvious distinction between Dirac MIN and Bosonic MIN is directly
caused by the differences between Fermi-Dirac and Bose-Einstein statistics.
The Dirac particles must obey the Pauli exclusion
principle and access only two quantum levels, while the Bosonic fields is
the infinite dimensional system.


\section{conclusions}

How the Unruh effect affects the MIN of Dirac and Bosonic fields
was investigated, and the following properties were found.

For Dirac fields, we choose the X-type states as initial state, and find
that: (i) the MIN $N(\rho_{A,I})$ decreases as the Unruh temperature increases,
while $N(\rho_{A,II})$ is contrary to it. However,
$\lim_{T\rightarrow\infty}N(\rho_{A,II})=\lim_{T\rightarrow\infty}N(\rho_{A,I})$, and
they always persists when $q_R\neq0$; (ii) both $N(\rho_{A,I})$ and
$N(\rho_{A,II})$ have a peculiar dynamics with a sudden change at
$T_{sc}$ provided $c_i$ is appropriately chosen. The $T_{sc}$ for
$N(\rho_{A,I})$ decreases as $c_i$ increases, while it is contrary
for $N(\rho_{A,II})$ case; and (iii) the MIN is more general than the quantum
nonlocality related to violation of Bell's inequalities. Besides, it is always equal
or larger than the geometric discord.

For Bosonic field, we take the Bell state, which is a special case of the X-type state, as initial state and
find that: (i) no matter for Alice-Rob modes or Alice-antiRob modes, both the MIN and geometric discord
eventually approach to zero when $T\rightarrow\infty$, that is to say, no ``quantum correlation" exists
in infinite acceleration limit; (ii) the choice of $q_R$ leads to two different dynamic behaviors of the MIN
and geometric discord. One is that the quantity firstly increases to an maximal value, then decays
monotonously, and finally approaches to zero, and the other is that the quantity degrades monotonously to
zero with the increase of $t$; and (iii) as a correlation, the MIN is always bigger than or equal to geometric
discord.

A distinctly distinguishable property of the MIN in non-inertial frames is that the MIN between Dirac
fields remains non-zero in the infinite acceleration limit, while for Bosonic fields it vanishes.
Which indicates that the MIN between Dirac fields is more robust than Bosonic fields case when they undergo
Unruh effect. Furthermore, for Dirac fields there is no fundamental difference of degradation (increase)
of MIN between Alice-Rob modes (Alice-antiRob modes) when we takes different $q_R$. However, for Bosonic fields
different $q_R$ will lead to different dynamic behaviors of its MIN. Different statistics for Dirac and Bosonic fields
directly result in these distinctions. Which suggests fundamental differences between infinite
and finite dimensional system MIN in relativistic settings.

Our discussion can be extended to the background of black hole by assuming that
Rob locates at the event horizon, while Alice freely falls into the black hole.
Therefore, by replacing the Unruh temperature with Hawking temperature, our result
shows that the MIN for Dirac fields persists in the infinite Hawking temperature limit,
while that for Bosonic fields vanishes. It is believed that further investigation
of relativistic quantum information may have implications for the problem of black hole
information loss.

\begin{acknowledgments}
This work was supported by the  National Natural Science Foundation
of China under Grant No. 11175065, 10935013; the National Basic
Research of China under Grant No. 2010CB833004; the SRFDP under
Grant No. 20114306110003; PCSIRT, No. IRT0964; the Hunan Provincial
Natural Science Foundation of China under Grant No 11JJ7001;  and
Construct Program of the National Key Discipline.
\end{acknowledgments}

\end{document}